\newcommand{\be}{\begin{equation}}
\newcommand{\bea}{\begin{eqnarray}}
\newcommand{\ee}{\end{equation}}
\newcommand{\eea}{\end{eqnarray}}
\newcommand{\nn}{\nonumber}

\newcommand{\qa}{\alpha}
\newcommand{\qb}{\beta}
\newcommand{\qg}{\gamma}
\newcommand{\qG}{\Gamma}
\newcommand{\qd}{\delta}

\newcommand{\qe}{\varepsilon}

\newcommand{\qth}{\theta}
\newcommand{\qy}{\theta}

\newcommand{\qk}{\kappa}
\newcommand{\ql}{\lambda}

\newcommand{\qr}{\rho}
\newcommand{\qs}{\sigma}

\newcommand{\qt}{\tau}

\newcommand{\qF}{\Phi}

\newcommand{\qj}{\psi}
\newcommand{\qJ}{\Psi}
\newcommand{\qo}{\omega}
\newcommand{\qO}{\Omega}

\newcommand{\tr}{{\rm tr}\,}

\newcommand{\dagg}{^{\dag}}

\newcommand{\fr}[2]{{\textstyle \frac{#1}{#2}}}

\newcommand{\EE}{\mathop{{\mathbb{E}}}}

\newcommand{\one}{{\mathbb 1}}

\newcommand{\bits}{ \{0,1\} }

\newcommand{\cA}{{\mathcal A}}
\newcommand{\cB}{{\mathcal B}}

\newcommand{\cE}{{\mathcal E}}
\newcommand{\cF}{{\mathcal F}}

\newcommand{\cH}{{\mathcal H}}
\newcommand{\cI}{{\mathcal I}}

\newcommand{\cK}{{\mathcal K}}

\newcommand{\cM}{{\mathcal M}}

\newcommand{\cO}{{\mathcal O}}
\newcommand{\cP}{{\mathcal P}}
\newcommand{\cQ}{{\mathcal Q}}

\newcommand{\cS}{{\mathcal S}}
\newcommand{\cT}{{\mathcal T}}

\newcommand{\cX}{{\mathcal X}}

\newcommand{\pr}{{\rm Pr}}

\newcommand{\isdef}{\stackrel{\rm def}{=}}

\newcommand{\ket}[1]{| #1 \rangle}
\newcommand{\bra}[1]{\langle #1 |}

\newtheorem{theorem}{Theorem}

\newtheorem{lemma}{Lemma}
\newtheorem{definition}{Definition}

\documentclass[11pt]{article}
\usepackage{graphics}
\usepackage{graphicx}
\usepackage{url}
\usepackage{amsmath}
\usepackage{amssymb}
\usepackage{amsrefs}
\usepackage{bbold}
\usepackage{a4wide}
\usepackage{enumitem}

\begin{document}

\setlength{\parindent}{0mm}

\title{Qubit-based Unclonable Encryption with Key Recycling}

\author{Daan Leermakers, Boris \v{S}kori\'{c}}

\date{ }

\maketitle

\begin{abstract}
\noindent 
We re-visit Unclonable Encryption as introduced by Gottesman in~2003 \cite{uncl}.
We look at the combination of Unclonable Encryption and Key Recycling,
while aiming  for low communication complexity and high rate.

\noindent
We introduce a qubit-based prepare-and-measure Unclonable Encryption scheme with re-usable keys. 
Our scheme consists of a single transmission by Alice and a single classical feedback bit from Bob.
The transmission from Alice to Bob consists entirely of qubits. 
The rate, defined as the message length divided by the number of qubits,
is higher than what can be achieved using Gottesman's scheme~\cite{uncl}.
We provide a security proof based on the diamond norm distance, taking noise into account.
\end{abstract}

\section{Introduction}

\subsection{Doing better than One-Time Pad encryption}
\label{sec:OTP}

Classically, the best confidentiality guarantee is provided by One-Time Pad (OTP) encryption.
If Alice and Bob share a uniform $n$-bit secret key, they can exchange an $n$-bit message with information-theoretic security. 
In the classical setting Eve is able to save a copy of the ciphertext. 
For the message to remain secure in the future, two conditions must be met:
\begin{enumerate}[leftmargin=5mm,itemsep=0mm]
\item The key is used only once.
\item The key is never revealed.
\end{enumerate}
If a quantum channel is available, these conditions can both be relaxed. 
(i) Quantum Key Recycling (QKR) 
\cite{BBB82,FehrSalvail2017,QKR_noise} schemes provide a way of re-using encryption keys.
(ii) Unclonable Encryption (UE) \cite{uncl} guarantees that a message remains secure even if the keys leak at some time in the future. 


In this paper we introduce a sheme that achieve both the key recycling and UE properties, and we explicitly prove 
that this can be achieved with low communication complexity. 
Our scheme acts only on individual qubits with simple prepare-and-measure operations.

\subsection{Quantum Key Recycling}
\label{sec:QKR}
The most famous use of a quantum channel in the context of cryptography is Quantum Key Distribution (QKD). 
First proposed in 1984 \cite{BB84}, QKD allows Alice and Bob to extend a small key, used for authentication, 
to a longer key in an information-theoretically secure way. 
Combined with classical OTP encryption this lets Alice and Bob exchange messages with theoretically unconditional security. 
The QKD field has received a large amount of attention, resulting in QKD schemes
that discard fewer qubits, various advanced proof techniques, improved noise tolerance, improved rates,
use of EPR pairs, higher-dimensional quantum systems etc.
\cite{Ekert91,Bruss1998,GottPres2001,SP2000,LoChauArdehali2004,RennerThesis,KGR2005,BHLMO2005,SYK2014,TL2017}.
Much less known is that the concept of QKR was proposed two years before QKD \cite{BBB82}.
QKR allows for the re-use of the secret encoding key when no disturbance is detected.
QKD and QKR have a lot in common.
(i)
They both encode classical data in quantum states, in a basis that is not a priori known to Eve.
(ii)
They rely on the no-cloning theorem \cite{WoottersZurek} to guarantee that without disturbing the quantum state, 
Eve can not gain information about the classical payload or about the basis. 

The security of QKD has been well understood for a long time \citelist{e.g. \cite{Bruss1998} \cite{SP2000}\cite{RennerThesis}\cite{TL2017}}, while a security proof for qubit-based QKR has been provided fairly recently \cite{FehrSalvail2017}. 
A cipher with near optimal rate using high-dimensional qudits was introduced in 2005 \cite{DBPS2014}. Unfortunately, their method requires a quantum computer to perform encryption and decryption. In 2017, a way of doing authentication (and encryption) of quantum states with Key Recycling was proposed \cite{Portmann2017}. However this work did not lead to a prepare-and-measure variant. 

The main advantage of QKR over QKD+OTP is reduced round complexity:
QKR needs only two rounds.
After the communication from Alice to Bob, only a single bit of authenticated information needs to be sent back from Bob to Alice. 
Recently, it was shown that QKR over a noisy quantum channel can achieve the same communication rate as QKD
(in terms of message bits per qubit)
even if Alice sends only qubits \cite{SilentBob};
a further reduction of the total amount of communicated data.

\subsection{Unclonable Encryption}
\label{sec:UE}

In 2003, D.\,Gottesman introduced a scheme called {\em Unclonable Encryption}\footnote{
This is slightly different from the unclonability notion of Broadbent and Lord \cite{BroadbentLord}
which considers two collaborating parties who both wish to recover the plaintext.
}
(UE) \cite{uncl} where the message 
remains secure even if the encryption keys leak at a later time 
(provided that no disturbance is detected).
His work was motivated by the fact that on the one hand many protocols require keys to be deleted,
but on the other hand permanent deletion of data from nonvolatile memory is a nontrivial task.
In this light it is prudent to assume that all key material which is not {\em immediately} discarded
is in danger of becoming public in the future;
hence the UE security notion demands that 
the message stays safe even if all this key material is made public after Alice and Bob
decide that they detected no disturbance.
(In case disturbance is detected, the keys have to remain secret forever or permanently destroyed.)

Gottesman remarked on the close relationship between UE and QKD, and in fact constructed a QKD variant from~UE. 
The revealing of the basis choices in QKD is equivalent to revealing keys in~UE.
It is interesting to note that Gottesman's UE construction allows partial re-use of keys.
However, it still expends one bit of key material per qubit sent.
In the current paper we introduce qubit-based
UE without key expenditure.

Since QKR sends a message directly instead of establishing a key for later use, 
QKR protocols are natural candidates to have the UE property.
In the case of noiseless quantum channels,
the high-dimensional encryption scheme \cite{DBPS2014} and
the qubit-based schemes \cite{FehrSalvail2017,QKR_noise} seem to have UE;
for noisy channels \cite{QKR_noise} with modified parameters
also seems to have UE.
However,
none of these conjectures have been explicitly stated or proven, 
which is a shame since resilience against key leakage is an interesting security feature. The QKR protocol where Alice sends only qubits \cite{SilentBob} is clearly not unclonable, due to the fact that single-use keys are stored at the end of each round.

\section{Contributions}
\label{sec:contrib}

We investigate the possibility of constructing an Unclonable Encryption scheme with recyclable keys,
while aiming for low communication complexity and high rate. 

We consider the following setting.
Alice and Bob have a reservoir of shared key material.
Alice sends data to Bob in $N$ chunks. 
Each chunk individually is tested by Bob for consistency (sufficiently low noise and valid MAC). 
In case of {\tt reject} they have to access new key material from the reservoir.
In case of {\tt accept}, Alice and Bob re-use their key material; this may be done either 
by keeping keys unchanged or by re-computing keys without accessing the reservoir.
If the $N$'th round was an {\tt accept}, all keys of round $N$ are assumed to become public.
\begin{itemize}[leftmargin=4mm,itemsep=0mm]
\item
We define the Key Recycling (KR) and Unclonable Encryption (UE) properties in terms
of the diamond norm. For this definition
we show a relation between KR and UE:
If a KR scheme re-uses all its long-term secrets in unchanged form upon
{\tt accept}, then it also has the UE property.
\item
We introduce {\tt KRUE}, a qubit-based prepare-and-measure scheme that satisfies KR and UE.
(Upon {\tt accept} some of the keys are kept and some are updated.)
Alice sends a single transmission, which consists entirely of qubits. 
Bob responds with a single classical feedback bit.
We provide a security proof by upper bounding the diamond distance between the protocol and 
its idealized functionality.
In particular, we use a reduction to the diamond distance that is associated with the security of QKD \cite{RennerThesis}.
In the case of a noiseless channel this reduction is almost immediate, without involving any inequalities.
For a noisy quantum channel {\tt KRUE}'s asymptotic communication rate equals the asymptotic QKD rate minus $h(\qb)$,
where $\qb$ is the bit error rate of the quantum channel and $h$ is the binary entropy function.
The difference in rate is caused by the fact that part of the message in {\tt KRUE} is a new mask
for protecting a syndrome in the next round.
\item
Next we introduce {\tt KRUE}$^*$, a stripped down version of {\tt KRUE} which needs an external mechanism for
transporting the next-round mask.
For the external mechanism we propose a QKR scheme \cite{SilentBob}.
This has the advantage that Alice still needs only a single transmission which consists entirely of qubits.
The rate of the combined scheme QKR+{\tt KRUE}$^*$ is higher than the rate of {\tt KRUE}
but lower than the QKD rate.
Also, if one would combine Gottesman's scheme with a secure key update mechanism in order to get the KR property,
the rate of that combination would be half the rate of QKR+{\tt KRUE}$^*$.
\end{itemize}

The outline of the paper is as follows.
After introducing notation and preliminaries in Section~\ref{sec:prelim}, 
we introduce the attacker model and security definition in Section~\ref{sec:secdefattackmodel}.  
We then introduce the {\tt KRUE} protocol (Section~\ref{sec:embedded}) 
and provide its security proof (Section \ref{sec:secproof}). 
In Section~\ref{sec:combine} we introduce {\tt KRUE}$^*$ and discuss its composition with
QKD and QKR as external mechanisms to transport the new mask.
Finally, in Section \ref{sec:tradeoff} we compare our schemes to existing qubit-based alternatives.

\section{Preliminaries}
\label{sec:prelim}

\subsection{Notation and terminology}
\label{sec:notation}

Classical Random Variables are denoted with capital letters, and their realisations
with lowercase letters. 
The expectation with respect to $X$ is denoted as 
$\EE_x f(x)=\sum_{x\in\cX}\pr[X=x]f(x)$.
For the $\ell$ most significant bits of the string $s$ we write $s[:$$\ell]$. 
The Hamming weight of a string $s$ is denoted as $|s|$.
The notation `$\log$' stands for the logarithm with base~2.
The notation $h$ stands for the binary entropy function $h(p)=p\log\fr1p+(1-p)\log\fr1{1-p}$.
Sometimes we write $h(p_1,\ldots,p_k)$ meaning $\sum_{i=1}^k p_i\log\fr1{p_i}$.
Bitwise XOR of binary strings is written as `$\oplus$'.
The Kronecker delta is denoted as $\qd_{ab}$.
We will speak about
`the bit error rate $\qb$ of a quantum channel'.
This is defined as the probability that a classical bit $x$, sent by Alice embedded in a qubit,
arrives at Bob's side as the flipped value $\bar x$. 
A linear error-correcting code with a $\ell \times n$ generator matrix 
$G$ can always be written in systematic form, $G = (\one_\ell | \qG)$, where the $\ell \times (n-\ell)$ matrix 
$\qG$ contains the checksum relations. 
For message $p\in\bits^\ell$,
the codeword $c_p = p\cdot G$ then has $p$ as its first $\ell$ bits, followed by $n-\ell$ redundancy bits.

For quantum states we use Dirac notation.
A qubit with classical bit $x$ encoded in basis $b$ is written as $\ket{\qj^b_x}$. 
The set of bases is~$\cB$.
In case of BB84 states we have $\cB=\{x,z\}$; in case of 6-state encoding
$\cB=\{x,y,z\}$.
The notation `tr' stands for trace.
Let $A$ have eigenvalues~$\ql_i$. 
The 1-norm of $A$ is written as $\|A\|_1=\tr\sqrt{A\dagg A}=\sum_i|\ql_i|$. 
States with non-italic label `A', `B' and `E' indicate the subsystem of Alice/Bob/Eve.

Consider classical variables $X,Y$ and a quantum system under Eve's control that depends on $X$ and~$Y$. 
The combined classical-quantum state is $\qr^{XY \rm E}=\EE_{xy} \ket{xy}\bra{xy} \otimes \qr^{\rm E}_{xy}$. 
The state of a sub-system is obtained by tracing out all the other subspaces, 
e.g. $\qr^{ Y \rm E}={\rm tr}_X \qr^{XY\rm E}=\EE_y \ket y\bra y\otimes\qr^{\rm E}_y$, with $\qr^{\rm E}_y=\EE_x\qr^{\rm E}_{xy}$.
The fully mixed state on $\cH_A$ is denoted as~$\chi^A$.
We also use the notation $\mu^X=\EE_x \ket x\bra x$ for a classical variable $X$ that is not necessarily uniform.

We write $\cS(\cH)$ to denote the space of density matrices on Hilbert space $\cH$, 
i.e.\,positive semi-definite operators acting on $\cH$. 
Any quantum channel can be described by a completely positive trace-preserving (CPTP) map 
$\cE: {\cS({\cH_{\rm A}})} \rightarrow {\cS(\cH_{\rm B})}$ that transforms a mixed state $\rho^{\rm A}$ 
to $\rho^{\rm B}$: $\cE(\rho^{\rm A}) = \rho^{\rm B}$.
For a map $\cE: S(\cH_{\rm A}) \rightarrow S(\cH_{\rm B})$, the notation $\cE(\rho^{\rm AC})$ stands for 
$(\cE \otimes \one_C) (\rho^{\rm AC})$, 
i.e.~$\cE$ acts only on the ‘A’ subsystem. 
Applying a map $\cE_1$ and then $\cE_2$ is written as the combined map $\cE_2\circ\cE_1$.
The diamond norm of $\cE$ is defined as $\| \cE \|_\diamond = \frac12 \sup_{\rho^{\rm AC} \in \cS( \cH_{\rm AC})} \| \cE(\rho^{\rm AC})\|_1$ with ${\cH_{\rm C}}$ an auxiliary system that can be considered to be of the same dimension as $\cH_{\rm A}$. 
The diamond norm $\|\cE-\cE'\|_\diamond$ can be used to
bound the probability of distinguishing two CPTP maps $\cE$ and $\cE'$ given that the process is observed once. 
The maximum probability of a correct guess is $\frac12 + \frac14 \| \cE - \cE' \|_\diamond$. 
In quantum cryptography, one proof technique considers Alice and Bob performing actions on noisy EPR pairs.
These actions are described by a CPTP map $\cE$ acting on the input EPR states and outputting classical 
outputs for Alice and Bob, and correlated quantum side information for Eve.
The security of such a protocol is quantified by the diamond norm between the actual map 
$\cE$ and an idealised map $\cF$ which produces perfectly behaving outputs (e.g. perfectly secret QKD keys). 
When $\| \cE - \cF \|_\diamond \leq \qe$ 
we can consider $\cE$ to behave ideally except with probability $\qe$; 
this security metric is {\em composable} with other (sub-)protocols \cite{TL2017}.

A family of hash functions $H=\{h:\cX\to\cT  \}$ is called pairwise independent 
(a.k.a.\,2--independent or strongly universal)
\cite{WegmanCarter1981} 
if for all distinct pairs $x,x'\in\cX$
and all pairs $y,y'\in\cT$ it holds that
$\pr_{h\in H}[h(x)=y \wedge  h(x')=y']=|\cT|^{-2}$.
Here the probability is over random~$h\in H$.
We define the rate of a quantum communication protocol as the number of message bits communicated per sent qubit.

\subsection{Post-selection}
\label{sec:post-selection}

For protocols that are invariant under permutation of their inputs it has been shown \cite{CKR2009} that security 
against collective attacks (same attack applied to each qubit individually)
implies security against general attacks, at the cost of extra privacy amplification. 
Let $\cE$ be a protocol that acts on $S(\cH_{\rm AB}^{\otimes n})$
and let $\cF$ be the perfect functionality of that protocol.
If for all input permutations $\pi$ there exists a map $\cK_\pi$ on the output such that $\cE \circ \pi = \cK_\pi \circ \cE$, then
\bea
\| \cE -\cF \|_\diamond &\leq& (n+1)^{d^2-1} \max_{\qs \in S(\cH_{\rm ABE})} \Big\| (\cE - \cF)  ( \qs^{\otimes n})\Big\|_1  
\label{eq:post-selection}
\eea
where $d$ is the dimension of the $\cH_{\rm AB}$ space. ($d=4$ for qubits).
The product form $\qs^{\otimes n}$ greatly simplifies the security analysis:
now it suffices to prove security against `collective' attacks,
and to pay a price $2(d^2-1)\log(n+1)$ in the amount of privacy amplification, which is negligible compared to $n$.

\subsection{Noise symmetrisation with random Pauli operators}
\label{sec:randomPaulis}

In \cite{RennerThesis} it was shown that for $n$-EPR states in factorised form,
as obtained from e.g. Post-selection, a further simplification is possible.
For each individual qubit $j$, Alice and Bob apply the Pauli operation $\qs_{\qa_j}$ to their half
of the EPR pair, with $\qa_j\in\{0,1,2,3\}$ random and public;
then they forget~$\qa$.
The upshot is that Eve's state (the purification of the Alice+Bob system) is simplified to
the $4\times4$ diagonal matrix Diag$(1-\fr32\qg, \fr\qg2, \fr\qg2, \fr\qg2)$.
Only one parameter is left over, the bit error probability~$\qg$ caused by Eve.
This symmetrisation trick is allowed when the {\em statistics} of the variables in the protocol
is invariant under the Pauli operations.

\section{Attacker model and security definitions}
\label{sec:secdefattackmodel}

\subsection{Attacker model}
\label{sec:attackmodel}

We work in same setting as Gottesman \cite{uncl}, as discussed in Section~\ref{sec:UE}. 
We distinguish between on the one hand long-term secrets 
and on the other hand short-term secrets. 
A variable is considered short-term only if it is
created\footnote{
Performing a measurement on a quantum state is also considered to `create' a classical variable.
} 
and immediately operated upon locally (without waiting for incoming communication),
and then deleted. 
All other variables are long-term.
(An example of a short-term variable is a nonce that is generated, immediately used a function evaluation and then deleted.
On the other hand, all keys that are stored between protocol rounds are long-term.)

We consider two world views.
\begin{itemize}[leftmargin=4mm,itemsep=0mm]
\item 
{\bf World1}.
All secrets can be kept confidential indefinitely or destroyed.
\item
{\bf World2}.
Long-term secrets are in danger of leaking at some point in time.
\end{itemize}
There are several motivations for entertaining the second world view.
(a) It is difficult to permanently erase data from nonvolatile memory.
(b) Whereas everyone understands the necessity of keeping message content confidential,
it is not easy to guarantee that protocol implementations (and users) handle the keys with the same care as the messages.

QKR protocols are typically designed to be secure in world1. 
In this paper we prove security guarantees that additionally hold in world2.
One way of phrasing this is to say that we add `user-proofness' to QKR.

Alice sends data to Bob in $N$ chunks. 
We refer to the sending of one chunk as a `round'.\footnote{
One data transmission will be called a {\em pass}. A round consists of multiple passes. 
} 
In each round Bob tells Alice if he noticed a disturbance (`{\tt reject}')
or not (`{\tt accept}').
In case of {\tt reject} they are {\em alarmed} and they know that they must take special care to protect the keys
of this round indefinitely
(i.e.~a fallback to {\em World1} security).
Crucially, {\em we assume that a key theft occurring before the end of round $N$
is immediately noticed by Alice and/or Bob}.
Without this assumption 
it would be impossible to do Key Recycling in a meaningful way.
We allow all keys to become public after round~$N$.

The rest of the attacker model consists of the standard assumptions:
no information, other than specified above, leaks from the labs of Alice and Bob; 
there are no side-channel attacks;
Eve has unlimited (quantum) resources; 
all noise on the quantum channel is considered to be caused by Eve.

\subsection{Security definitions}
\label{sec:secdef}

Let $\qr^{MK\tilde KT \rm E}=\qr_{\tt accept}^{MK\tilde KT \rm E}+\qr_{\tt reject}^{MK\tilde KT \rm E}$ 
be the state after execution of a quantum encryption protocol,
where $M$ is the classical message, $K$ stands for all the keys (and other long-term secrets), $\tilde K$ the updated keys, 
$T$ the transcript observed by Eve,
and E Eve's quantum side information.
The two parts, associated with outcomes {\tt accept} and {\tt reject} respectively, are sub-normalized. 
In some existing QKR protocols, e.g. \cite{FehrSalvail2017}, the keys remain unchanged ($\tilde K=K$) in case of {\tt accept},
whereas in other protocols, e.g. \cite{SilentBob}, there is always a key update. 
Alice and Bob share a `reservoir' of key material from which key refreshes are done in case of {\tt reject}.

The {\it Encryption} property ({\bf ENC}) is defined as follows.
\begin{definition}
\label{def:ENC}
An encryption scheme with output $\qr^{MKT\rm E}$ is called $\qe$-encrypting ($\qe$-ENC) if it satisfies
\be
	 \|\qr^{MT\rm E}-\qr^M\otimes\qr^{T\rm E}\|_1\leq\qe.
\ee
\end{definition}
Furthermore we will work with the following definitions.
\begin{definition}
\label{def:KR}
A scheme with output $\qr^{MK\tilde KT\rm E}$ is called $\qe$-recycling ($\qe$-KR) if 
(i) the reservoir is not accessed for creating the updated keys $\tilde K_{\tt accept}$
and (ii) it satisfies
\be
	\| \qr^{M\tilde KT\rm E}- \qr^{\tilde K}\otimes\qr^{MT\rm E} \|_1  \leq\qe.
\label{defKR}
\ee
\end{definition}

\begin{definition}
\label{def:UE}
A scheme with output $\qr^{MK\tilde KT\rm E}$ is called $\qe$-unclonable ($\qe$-UE) if it satisfies
\be
	 \| \qr_{\tt accept}^{MK\tilde KT\rm E}-\qr^M\otimes\qr_{\tt accept}^{K\tilde KT\rm E}  \|_1 \leq\qe.
\ee
\end{definition}

Note that other definitions exist. 
For instance, \cite{FehrSalvail2017} has a recycling definition that allows Eve to have partial information
about one of the keys (the measurement basis), as long as the min-entropy is high enough.
Our preference for the above KR and UE definitions stems from 
(i) the fact that it allows for a unified treatment of all the keys;
(ii) compatibility with the proof technique of \cite{RennerThesis,CKR2009},
which makes it possible to prove security of high-rate schemes. 

Furthermore our KR definition is compatible with~\cite{DBPS2014}.
Also note that our KR and UE do not automatically imply ENC. 
The ENC property has to be considered as a separate requirement.
For the combination of ENC and KR we have the following two lemmas.
\begin{lemma}
\label{lemma:singlenorm}
\be
	\| \qr^{M\tilde KT\rm E}- \qr^M\otimes\qr^{\tilde K}\otimes\qr^{T\rm E} \|_1 \leq \qe
	\quad\implies\quad
	\qe\mbox{-ENC} \;\;\wedge\;\; 2\qe\mbox{-KR}
\label{impliesENCKR}
\ee
\end{lemma}
\underline{\it Proof}.
Taking the lhs of (\ref{impliesENCKR}) and tracing over $\tilde K$ yields $\qe$-ENC.
Furthermore, using the triangle inequality we write
$\|\qr^{M\tilde KT\rm E}-\qr^{\tilde K}\otimes\qr^{MT\rm E}\|_1$ 
$\leq \| \qr^{M\tilde KT\rm E}-\qr^M\otimes\qr^{\tilde K}\otimes\qr^{T\rm E}  \|_1  $
$+\|  \qr^M\otimes\qr^{\tilde K}\otimes\qr^{T\rm E}-\qr^{\tilde K}\otimes\qr^{MT\rm E}  \|_1$.
Both terms individually are bounded by $\qe$ by the lhs of (\ref{impliesENCKR});
the first term directly, the second term after taking the $\tilde K$-trace. This proves $2\qe$-KR.
\hfill$\square$

\begin{lemma}
\label{lemma:impliesUE}
\be
	(\tilde K_{\tt accept}=K) \;\;\wedge\;\; \qe_1\mbox{-ENC} \;\;\wedge\;\; \qe_2\mbox{-KR}
	\quad\implies\quad (\qe_1+\qe_2)\mbox{-UE}.
\ee
\end{lemma}
\underline{\it Proof}.
With $\tilde K_{\rm accept}=K$ we have 
$\| \qr_{\tt accept}^{MK\tilde KT\rm E}-\qr^M\otimes\qr_{\tt accept}^{K\tilde KT\rm E}  \|_1$
$\leq \|  \qr_{\tt accept}^{MKT\rm E}- \qr^M\otimes\qr^K\otimes\qr_{\rm accept}^{T\rm E}  \|_1$
$+ \|  \qr^M\otimes\qr^K\otimes\qr_{\rm accept}^{T\rm E} - \qr^M\otimes\qr_{\tt accept}^{KT\rm E}  \|_1$.
The first term is bounded by taking the trace over $K$ and using $\qe_1$-ENC.
For the second we take the trace over $M$, yielding
$ \|  \qr_{\tt accept}^{KT\rm E} - \qr^K\otimes\qr_{\rm accept}^{T\rm E} \|_1$.
This expression is bounded by $\qe_2$, which is seen by taking the $M$-trace of (\ref{defKR}).
\hfill$\square$

\vskip3mm

Lemma~\ref{lemma:singlenorm} allows us to prove both ENC and KR by upperbounding a single quantity.
Lemma~\ref{lemma:impliesUE} is an important statement:
any ENC scheme that upon {\tt accept} re-uses its keys {\em in unmodified form}
and satisfies KR is automatically a UE scheme.
It is interesting to note that \cite{FehrSalvail2017} has $\tilde K_{\tt accept}=K$ but does not
satisfy KR, whereas \cite{QKR_noise,SilentBob} satisfies KR but does not have $\tilde K_{\tt accept}=K$. 
By Theorem~4 in \cite{DBPS2014} and Lemma~\ref{lemma:impliesUE},
the {\em high-dimensional} scheme of Damg{\aa}rd et al.~\cite{DBPS2014} has the UE property.

\subsection{CPTP maps}
\label{sec:secCPTP}

We consider again the sequence of $N$ chunks.
The KR property must hold in the first $N-1$ rounds.
The ENC and UE property must hold in all rounds.
We write the statements of Section~\ref{sec:secdef} in terms of CPTP maps, 
to make contact with the proof technique of Section~\ref{sec:post-selection}.

The different nature of the KR and UE property forces us to introduce two different notations
for the CPTP map that is executed by Alice and Bob.
On the one hand, we write $\cE_{\rm KR}$ for one round of the protocol,
where at the end of the round the old keys (from the beginning of the round) are traced away.
On the other hand, we write $\cE_{\rm UE}$ for one round without such a tracing operation.
(The ENC property is not made explicit in this notation.)
The following condition implies that the above given security properties hold except with probability~$\qe$,
\be
	\forall_{j\in\{1,\ldots,N\}}\quad \left\|   \cE_{\rm UE}^{(j)}\circ\cE_{\rm KR}^{(j-1)}\circ\cdots\circ\cE_{\rm KR}^{(1)}  
	- \cF_{\rm UE}^{(j)}\circ\cF_{\rm KR}^{(j-1)}\circ\cdots\circ\cF_{\rm KR}^{(1)}
	\right\|_\diamond\leq \qe, 
\label{conditionallj}
\ee
where the superscript is the round index, and $\cF$ stands for the idealized version of the protocol.
We can arrive at a simplified statement using the following lemma.

\begin{lemma}
\label{lem:maps}
For any CPTP maps $\cA, \cA', \cB, \cB'$, it holds that
\bea
\| \cA \circ \cB - \cA' \circ \cB' \|_\diamond 
&\leq& \| \cA - \cA' \|_\diamond + \| \cB - \cB' \|_\diamond.
\eea
\end{lemma}
\underline {Proof}:
\bea
\| \cA \circ \cB - \cA' \circ \cB' \|_\diamond &=& \| \cA \circ \cB - \cA' \circ \cB' + \cA' \circ \cB - \cA' \circ \cB \|_\diamond
\\&\leq& \| (\cA - \cA') \circ \cB \|_\diamond + \| \cA'\circ(\cB - \cB') \|_\diamond
\\&\leq& \| \cA - \cA' \|_\diamond + \| \cB - \cB' \|_\diamond
\eea
where the last inequality holds because a CPTP map can never increase the trace distance.
\hfill $\square$

Using Lemma~\ref{lem:maps} it is easily seen that the following condition implies~(\ref{conditionallj}),
\be
	(N-1) \| \cE_{\rm KR}-\cF_{\rm KR} \|_\diamond +
	\| \cE_{\rm UE} - \cF_{\rm UE} \|_\diamond   \leq \qe.
\ee

It is therefore sufficient to upper bound the single-round quantities
$\| \cE_{\rm KR}-\cF_{\rm KR} \|_\diamond$ and
$\| \cE_{\rm UE} - \cF_{\rm UE} \|_\diamond$.
The ideal mapping $\cF$ is obtained from $\cE$ as follows.
From $\cE(\qr^{\rm ABE})$ one traces out those classical variables that are supposed to remain unknown to Eve,
and takes a tensor product with an isolated mixed state of these variables.
In the case of $\cE_{\rm KR}$ the relevant variables are the message $m$ and the next-round keys,
which we denote here as~$\tilde k$.
In the case of $\cE_{\rm UE}$ it is only the message, and only the {\tt accept} part of the mapping is relevant.
(Upon {\tt reject} the functionality of $\cE_{\rm UE}$ is ideal by definition.)
Hence we have 
\bea
	\| \cE_{\rm KR} - \cF_{\rm KR} \|_\diamond 
	&=& 
	\frac12 \sup_{\rho^{\rm ABE}} \Big\|  \cE_{\rm KR} (\rho^{\rm ABE}) 
	- \EE_{m \tilde k} \ket{m\tilde k}\bra{m\tilde k} \otimes \tr_{M \tilde K}\, \cE_{\rm KR} (\rho^{\rm ABE})\Big\|_1
	\\
	\| \cE_{\rm UE} - \cF_{\rm UE} \|_\diamond 
	&=& 
	\frac12 \sup_{\rho^{\rm ABE}} \Big\|  \cE^{\tt accept}_{\rm UE} (\rho^{\rm ABE}) - 
	\EE_m \ket{m}\bra{m} \otimes \tr_{M} \, \cE^{\tt accept}_{\rm UE} (\rho^{\rm ABE})\Big\|_1,
\label{EminFunclgen}
\eea
where in (\ref{EminFunclgen}) the {\tt reject} part vanishes as we have implicitly assumed that
$\cE$ has ideal functionality in the {\tt reject} case, i.e. ENC holds.


\section{The proposed scheme, {\tt KRUE}}
\label{sec:embedded}

We propose a qubit-based prepare-and-measure
scheme for Unclonable Encryption with Key Recycling.
We will refer to it as {\tt KRUE}.

\subsection{Pairwise independent hashing with easy inversion}
\label{sec:invertible_hash}

We will need the privacy amplification step to be easily computable in two directions. 
Unfortunately the code-based
construction due to Gottesman \cite{uncl} does not work with the proof technique of \cite{RennerThesis}, which
requires a family of universal hash functions.
We will be using a family of invertible functions $F: \bits^\nu \to \bits^\nu$ that has the collision properties of a pairwise independent hash function.
An easy way to construct such a family is to use multiplication in $GF(2^\nu)$.
Let $u \in GF(2^\nu)$ be randomly chosen. 
Define $F_u(x) = u \cdot x$, where the multiplication is in $GF(2^\nu)$. 
A pairwise independent family of hash functions $\qF$ from  $\bits^\nu$ to $\bits^\ell$, with $\ell \leq \nu$, is implemented by taking the $\ell$ 
most significant bits of $F_u(x)$. 
We denote this as
\be
	\qF_u(x) = F_u(x)[:\!\ell].
\ee
The inverse operation is as follows. Given $c\in\bits^\ell$, generate random $r\in\bits^{\nu-\ell}$ and output $F_u^{\rm inv}(c||r)$.
It obviously holds that $\qF_u( F_u^{\rm inv}(c||r) )=c$.
Computing an inverse in $GF(2^\nu)$ costs $O(\nu \log^2 \nu)$ operations \cite{Moenck73}.

\subsection{Protocol steps}
\label{sec:protocol}

Alice and Bob have agreed on a MAC function $\qG:\bits^\ql\times\bits^{\ell-\ql}\to\bits^\ql$, 
the function families $F$ and $\qF$
as discussed in Section \ref{sec:invertible_hash}, with $\nu=k$,
and a linear error correcting code 
which has encoding function ${\rm Enc}:\bits^k \to\bits^n$ in systematic form
and decoding ${\rm Dec}: \bits^n \to \bits^k$.

In round $j$, Alice wants to send a message $\mu_j\in\bits^{\ell-(n-k)-2\ql-1}$.
We will often drop the index $j$ for notational brevity. 
The key material shared between Alice and Bob consists of a mask $z \in \bits^\ell$, 
a MAC key $k_{\rm MAC} \in \bits^\ql$, a basis sequence $b \in \cB^n$, 
a MAC key $k_{\rm fb} \in \bits^\ql$ for the feedback bit, 
a one-time pad $k_{\rm OTP} \in \bits$ for the feedback bit, 
a key $u\in\bits^k$ for universal hashing  
and a key $e \in \bits^{n-k}$ to mask the redundancy bits. 
They have a reservoir of spare key material ($k_{\rm rej}$) from which
to refresh their keys in case of {\tt reject}.

\begin{figure}[h]
\begin{center}
\includegraphics[width=0.9\textwidth]{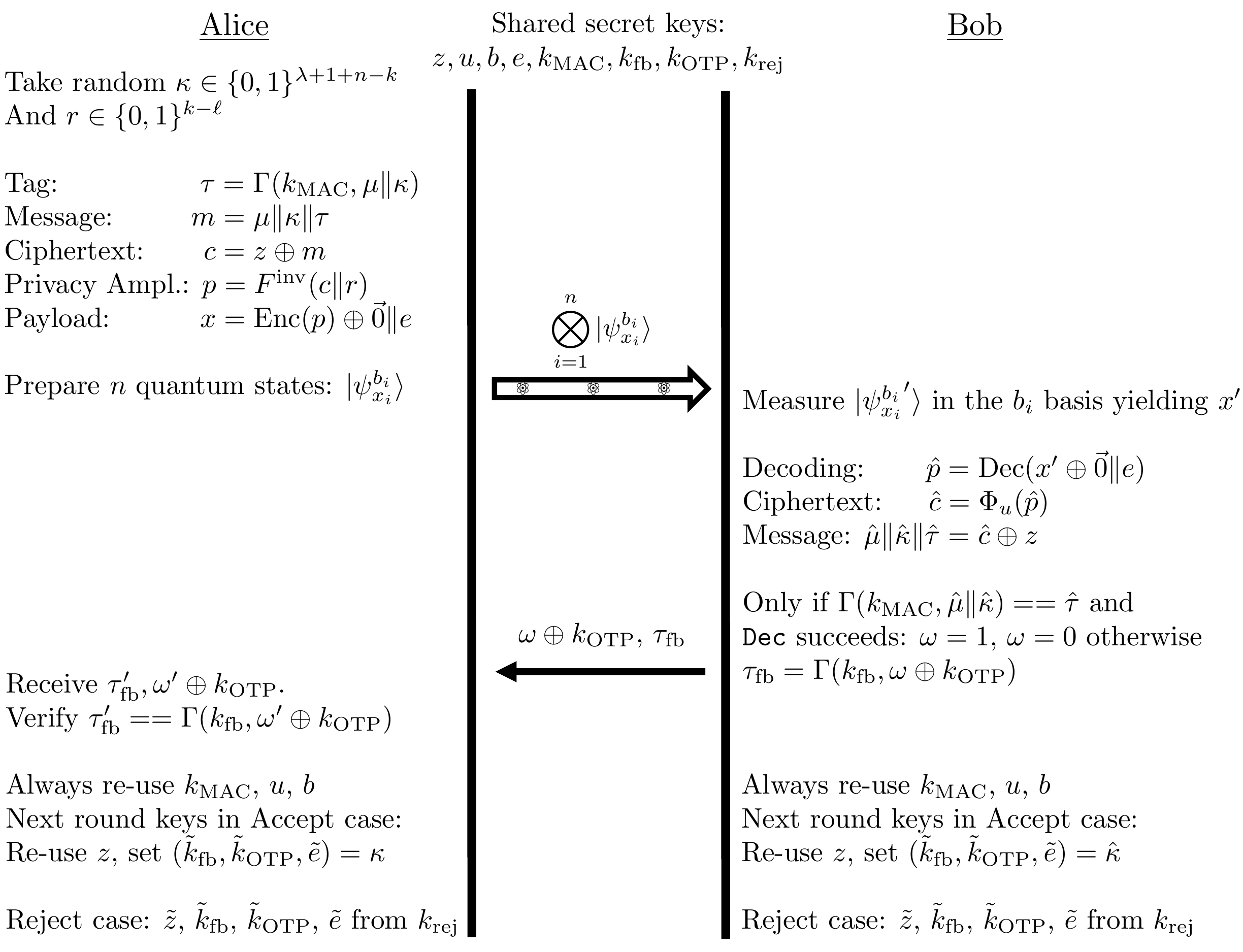}
\caption{\it Protocol steps of a single round.}
\label{fig:embedded}
\end{center}
\end{figure}

In each round
Alice and Bob perform the following steps (see Fig.\,\ref{fig:embedded}):

\underline{Encryption}: \\
Alice generates random strings  $\qk \in \bits^{\ql+1+n-k}$ and $r \in \bits^{k-\ell}$. 
She computes the authentication tag $\tau  = \qG(k_{\rm MAC}, \mu\| \qk)$, 
the augmented message $m = \mu \| \qk\|\tau$, the ciphertext $c=z\oplus m$, 
the reversed privacy amplification $p = F^{\rm inv}_u(c\|r) \in \bits^k$ 
and the qubit payload $x = {\rm Enc}(p) \oplus (\vec{0}^k\|e) \in \bits^{n}$.
She prepares $\ket\qJ = \bigotimes_{i=1}^n\ket{\qj^{b_i}_{x_i}}$ and sends it to Bob.

\underline{Decryption}:\\
Bob receives $\ket\qJ'$. 
He measures $\ket\qJ'$ in the basis $b$. 
The result is $x'\in\bits^n$. 
He decodes $\hat p = {\rm Dec}(x' \oplus(\vec{0}_k\|e))$.  
He computes $\hat c  =\Phi_u(\hat p)$ and $\hat \mu\|\hat \qk\| \hat \tau= \hat c \oplus z$.
He sets $\qo=1$ ({\tt accept}) if $\qG(k_{\rm MAC},\hat \mu\| \hat \qk)==\hat \qt$, otherwise $\qo=0$ ({\tt reject}). 
He computes $\tau_{\rm fb} = \qG(k_{\rm fb}, \qo \oplus k_{\rm OTP})$ and sends $\qo\oplus k_{\rm OTP}$ and $\tau_{\rm fb}$ to Alice.

\underline{Key Update}:\\
Alice and Bob perform the following actions (a tilde denotes the key for the next round): \\
In case of {\tt accept}: Re-use $b,u,k_{\rm MAC},z$. Set next round keys: ($\tilde k_{\rm fb}, \tilde k_{\rm OTP}, \tilde e) = \qk$.\\
In case of {\tt reject}: Re-use $b,u,k_{\rm MAC}$. Take fresh $\tilde z$, $\tilde k_{\rm fb}$, $\tilde k_{\rm OTP}$, $\tilde e$ from $k_{\rm rej}$.

After round~$N$, according to the attacker model, all keys from all rounds leak\footnote{
Optionally this leakage can be made part of the protocol, i.e. Alice and Bob publish the keys.
}
except for masks $z$ associated with {\tt reject} events. 
I.e.~what leaks is:
$b,u,k_{\rm MAC}$,
$\{ k_{\rm fb}^{(j)}, k_{\rm OTP}^{(j)}, e^{(j)}  \}_{j=1}^N$,
and if round $N$ was {\tt accept} also $z^{(N)}$.


{\it Remarks:}

\vspace{-1mm}

\begin{itemize}[leftmargin=4mm,itemsep=0mm]
\item
The augmented message $m$ contains the three keys $\tilde k_{\rm fb}, \tilde k_{\rm OTP}, \tilde e$
for the next round.
This means that qubits are `spent' in order to send something other than $\mu$, which reduces the communication rate.
Here the mask $e\in\bits^{n-k}$ for the redundancy bits
is the dominant part; its size is asymptotically $nh(\qb)$ bits, giving rise to a
rate penalty term $h(\qb)$ familiar from QKD.
\item
The {\tt accept}/{\tt reject} feedback bit is encrypted, which
temporarily prevents Eve from gaining information from `oracle' access to the feedback.
This allows us to re-use $b$ in unmodified form after {\tt accept}.
\item
Even in the case of known plaintext, from Eve's point of view the `payload' $x\in\bits^n$
in the state $\bigotimes_{i=1}^n\ket{\qj^{b_i}_{x_i}}$
is uniformly distributed. 
The $z$ masks $\ell$ bits; then appending $r$ increases that to $k$ bits; finally the $e$ masks the
 $n-k$ redundancy bits.
\end{itemize}

\subsection{EPR version of the  protocol}
\label{sec:EPR}

We will base the security proof on the EPR version of the protocol, making use of
Post-selection (Section~\ref{sec:post-selection}) and the random-Pauli noise symmetrisation technique
(Section~\ref{sec:randomPaulis}).

$n$ noisy singlet states are produced by an untrusted source, e.g.~Eve. 
One half of each EPR pair is sent to Alice, the other half to Bob. 
Alice and Bob apply the random Pauli operations as described in Section~\ref{sec:randomPaulis}.
Then Alice measures her qubits in the bases $b\in\cB^n$ resulting in a string~$s\in\bits^n$. 
Bob too measures his qubits in basis $b$, which yields $t\in\bits^n$. 
Alice computes $x$ as specified in Section~\ref{sec:protocol},
then computes $a=x\oplus s$ and sends $a$ to Bob over an authenticated classical channel.
Bob receives $a$, computes $x'=\bar t\oplus a$ and performs the decryption steps specified in Section~\ref{sec:protocol}.

We are allowed to use Post-selection because our protocol is invariant under permutation
of the EPR pairs.
A permutation re-arranges the noise in the observed strings $s$ and $t$ over the bit positions $\{1,\ldots,n\}$; 
however, the error correction step is insensitive to such a change.
The use of the noise symmetrisation technique is allowed because the statistics is invariant under
the Pauli operations. 
In the case of BB84 encoding and 6-state encoding, the Paulis cause bit flips in the string $x\in\bits^n$ in positions known to Alice and Bob,
which does not change the protocol in any essential way.\footnote{
In 8-state encoding \cite{SdV2017}, applying a Pauli changes the basis $b$ in a way known to Alice and Bob.
Again, this does not affect the security.
}
Security of the EPR version implies security of the prepare-and-measure protocol of Section~\ref{sec:protocol}.

\section{Security proof for the EPR verison of the protocol}
\label{sec:secproof}

\subsection{CPTP maps}
\label{sec:CPTP}

We now specify the exact form of the CPTP map which represents one round.
We start with $\cE_{\rm UE}$ and write $\cE_{\rm KR}=\cT_{\rm KR}\circ\cE_{\rm UE}$,
where $\cT_{\rm KR}$ is a partial trace operation.
The $\cE_{\rm UE}$ can be viewed as four consecutive maps: 
an initialization step $\cI$ where the input variables are prepared;
a measurement step~$\cM$; 
a post-processing step~$\cP$ representing all further computations; 
a partial trace step $\cT_{\rm UE}$ where all variables that are not part of the output or the transcript are traced away,
\be
	\cE_{\rm UE} = \cT_{\rm UE} \circ \cP \circ \cM \circ \cI.
\ee
The initialization merely appends the input variables,
\be
	\cI (\rho^{\rm ABE}) = \EE_{mbzue} \ket{mbzue}\bra{mbzue} \otimes \rho^{\rm ABE}.
\ee
Here $b,z,u,e$ are uniform, but $m$ not necessarily.
The measurement acts on the $b$-space and $\qr^{\rm ABE}$, outputting the strings $s,t$
and Eve's state $\qr^{\rm E}_{bst}$, which is correlated to the measurement basis $b$ and the outcomes $s,t$,
\be
	\cM (\ket b\bra b \otimes \qr^{\rm ABE}) = \EE_{st} \ket{bst}\bra{bst} \otimes \qr^{\rm E}_{bst}.
\ee
Here the distribution of $s$ and $t$ is governed by the i.i.d.~noise with noise parameter~$\qg$.
The marginals of $s$ and $t$ are uniform, 
while for all $j\in\{1,\ldots,n\}$ it holds that $\pr[s_j= t_j]=\qg$.

In the post-processing the flag $\qo$ is computed as a function of $s$ and $t$ which we will denote as
$\qy_{st}$.
Let $n\qb$ be the number of errors that can be corrected by the error-correcting code.
Then
\be
	\qy_{st}=\left\{ \begin{matrix}
	1 \mbox{ if }|\bar s\oplus t|\leq n\qb  \cr 
	0 \mbox{ if }|\bar s\oplus t| > n\qb  \end{matrix}
	\right. .
\ee
We will use the notation $P_{\rm corr}(n,\qb,\qg)$ for the probability of the event $\qy_{st}=1$.

\be
	P_{\rm corr}(n,\qb,\qg) \isdef  \EE_{st}\qy_{st}
	= \sum_{c=0}^{\lfloor n\qb\rfloor} {n\choose c}\qg^c(1-\qg)^{n-c}.
\label{defPcorr}
\ee
The result of applying $\cI,\cM,\cP$ is given by
\bea
	(\cP \circ \cM \circ \cI)(\qr^{\rm ABE}) 
	\!\! &=& \!\!\!\! 
	\EE_{mbzuest} \ket{mbzuest}\bra{mbzuest} \otimes \rho^{\rm E}_{bst} \otimes 
	\!\! \sum_{capx x' \qo\tilde z} \EE_{r} \ket{capxx'\qo \tilde zr}\bra{capxx'\qo \tilde z r}
	\nn\\&&
	\qd_{a,s\oplus x} \qd_{c,m\oplus z} \qd_{p,F_u^{\rm inv}(c\|r)} \qd_{x,p\| 
	[{\rm Red}(p) \oplus e]} \qd_{x', \bar t \oplus a}\qd_{\qo,\qth_{st}} 
	\big[ \qo\qd_{\tilde z z} +\frac{ \overline\qo }{2^{\ell}}\big].
\eea
Here $r$ is uniform and `Red$(p)$' stands for the redundancy bits appended to $p$ in the systematic-form ECC encoding Enc$(p)$. 
In the final step $\cT_{\rm UE}$ we trace away all variables that are not part of the transcript or the output: $s,t,c,p,x,x',r$.
These variables exist only temporarily and can be quickly discarded by Alice and Bob; they are never stored in
nonvolatile memory.
The $a$ and $\qo$ are observed by Eve as part of the communication.
(The $\qo$ in encrypted form, but the key is assumed to leak in the future.)
The $b,z,u,e$ are assumed to leak in the future and thus they have to be kept as part of the state.
We obtain\footnote{
Note that tracing out $u$ or $z\tilde z$ in (\ref{Euncl}) yields a state in which the $m$-subspace is completely decoupled 
from the rest of the Hilbert space. This shows that the scheme, when merely viewed as an encryption scheme,
protects $m$ unconditionally as soon as the adversary does not know~$u$ or $z\tilde z$.
}

\bea
	\cE_{\rm UE} (\qr^{\rm ABE}) 
	&=& 
	\EE_{mbzue} \sum_{a\tilde z \qo} \ket{mbzuea  \tilde z \qo}\bra{mbzuea  \tilde z \qo} \otimes  
	\EE_{st} \qr_{bst}^{\rm E} \sum_p 2^\ell \qd_{\qF_u(p), m \oplus z}
	\nn\\&& 
	2^{-k}\qd_{s\oplus a,p\|[{\rm Red}(p)\oplus e]}  \qd_{\qo,\qy_{st}}\Big[ \qo \qd_{\tilde z z} +{\overline\qo} 2^{-\ell}\Big].
\label{Euncl}
\eea
As discussed in Section~\ref{sec:secdefattackmodel}, only the {\tt accept} part (the $\qo=1$ part) of the idealized $\cF_{\rm UE}$
is relevant. This is obtained as 
$\cF^{\tt accept}_{\rm UE}(\qr^{\rm ABE})=\EE_m\ket m\bra m\otimes \tr_{\!M} \cE^{\rm accept}_{\rm UE}(\qr^{\rm ABE})$.
We get
\bea
	\cF^{\tt accept}_{\rm UE} (\qr^{\rm ABE}) 
	&=& \!\!\!
	\EE_{mbzue} \sum_{a\tilde z} \ket{mbzuea  \tilde z}\bra{mbzuea  \tilde z} \qd_{\tilde z z}
	\nn\\ &&
	\otimes  \EE_{st} \qr_{bst}^{\rm E}\qy_{st} \!\!\sum_p \! 2^{\ell-k}\qd_{s\oplus a,p\|[{\rm Red}(p)\oplus e]} 
	\!\EE_{m'} \qd_{\qF_u(p), m' \oplus z}.
\label{Funcl}	
\eea
Note that this expression is sub-normalized; its trace equals $P_{\rm corr}$.
We write
\bea
	&&
	(\cE^{\rm accept}_{\rm UE}- \cF^{\rm accept}_{\rm UE})(\qr^{\rm ABE}) =	 
	\EE_{mbzue} \sum_{a\tilde z} \ket{mbzuea  \tilde z}\bra{mbzuea  \tilde z} \qd_{\tilde z z}
	\nn\\ && \quad\quad\quad
	\otimes  \EE_{st} \qr_{bst}^{\rm E}\qy_{st} \sum_p2^{\ell-k}\qd_{s\oplus a,p\|[{\rm Red}(p)\oplus e]} 
	[\qd_{\qF_u(p), m \oplus z}-\EE_{m'} \qd_{\qF_u(p), m' \oplus z}].
\label{diffUncl}
\eea
For the description of $\cE_{\rm KR}$ we have to take (\ref{Euncl}) and trace out $z,e,\qo$.
\bea
	\cE_{\rm KR} (\qr^{\rm ABE})
	&=& 
	\EE_{mbu}2^{-n}\sum_{a\tilde z}\ket{mbua\tilde z}\bra{mbua\tilde z} \otimes  \EE_{st} \rho_{b s t}^{\rm E}  
	\Big[ \qth_{st} \qd_{\Phi_u((s\oplus a)[:k]), m \oplus \tilde z} +2^{-\ell}\overline{\qy_{st}}\Big].
\eea
The ideal functionality $\cF_{\rm KR}$ has $m,b,u,\tilde z$ decoupled from the rest of the system.
We have
$\cF_{\rm KR}(\qr^{\rm ABE})=\EE_{mbu}2^{-\ell}\sum_{\tilde z}\ket{mbu\tilde z}\bra{mbu\tilde z}\otimes\tr_{\!MBU\tilde Z}\cE_{\rm KR}(\qr^{\rm ABE})$,
which yields
\be
	\cF_{\rm KR} (\qr^{\rm ABE}) = \EE_{mbu}2^{-n-\ell}\sum_{a\tilde z}\ket{mbua\tilde z}\bra{mbua\tilde z} \otimes
	\EE_{st}\EE_{b'}\qr^{\rm E}_{b'st}.
\ee
Note that $\EE_{st}\EE_{b'}\qr^{\rm E}_{b'st}=\qr^{\rm E}$. 

\begin{lemma}
\label{lem:basisind}
Let $\qr^{\rm ABE}$ denote the purification of a $4^n$-dimensional state $\qr^{\rm AB}$. 
Let $b\in\cB^n$ be a qubit-wise orthonormal basis. 
It holds that
$\qr^{\rm E}_b = \qr^{\rm E}$.
\end{lemma}
\underline{\it Proof:}
Let $P^{\rm A}_{bs}$ denote a projection operator
on subsystem `$\rm A$' corresponding to a measurement
in basis $b$ with outcome $s\in\bits^n$.
We have 
$\qr_b^{\rm E} \isdef \EE_{st}\qr^{\rm E}_{bst}$
$= \sum_{st} \tr_{\rm AB} (P^{\rm A}_{bs} \otimes P^{\rm B}_{bt}\otimes\one)  \rho^{\rm ABE}$ 
$=\tr_{\rm AB} ([\sum_s P^{\rm A}_{bs}] \otimes [\sum_t P^{\rm B}_{bt}]\otimes\one)  \rho^{\rm ABE}$
$= \rho^{\rm E}$.
We use the fact that $\sum_s P^{\rm A}_{bs}=\one$ and $\sum_t P^{\rm B}_{bt}=\one$ for any $b$.
\hfill $\square$

Lemma~\ref{lem:basisind} allows us to write
\be
	(\cE_{\rm KR}-\cF_{\rm KR}) (\qr^{\rm ABE}) =
	\EE_{mbu}2^{-n-\ell}\sum_{a\tilde z}\ket{mbua\tilde z}\bra{mbua\tilde z} \otimes  \EE_{st} \rho_{b s t}^{\rm E}
	\qy_{st}[2^\ell\qd_{\Phi_u((s\oplus a)[:k]), m \oplus \tilde z}  -1].
\label{diffKR}
\ee

\subsection{Intermezzo: QKD asymptotics}
\label{sec:secQKD}

In Appendix \ref{app:QKD}, we consider a version of QKD where privacy amplification is implemented as
in Section~\ref{sec:protocol}, and the syndrome is sent to Bob in OTP'ed form;
we show that this leads to a bound of the form
\be
	\| \cE_{\rm QKD}-\cF_{\rm QKD} \|_\diamond \leq \fr12 \EE_{mbu}\frac1{2^{n+\ell}}\sum_{ac}
	\left\|  \EE_{st}\qr^{\rm E}_{bst}\qy_{st}2^\ell
	[\qd_{c,m\oplus\qF_u(a\oplus s)} - \EE_{m'}\qd_{c,m'\oplus\qF_u(a\oplus s)}] \right\|_1,
\label{refQKD}
\ee
which after some algebra gives rise to
\be
	\| \cE_{\rm QKD} - \cF_{\rm QKD} \|_\diamond 
	\leq \min \Big(P_{\rm corr} , 
	\frac12 \EE_{b} \tr \sqrt{ 2^\ell\EE_{ss'}\qd_{s s'} \rho^{\rm E}_{bs}\rho^{\rm E}_{bs'} 
	 }\Big),
\label{refQKD2}
\ee
and that from (\ref{refQKD2}) the well known asymptotic QKD rates is obtained:
$1-2h(\qb)$ for BB84 and $1-h(1-\frac{3\qb}{2},\frac{\qb}{2},\frac{\qb}{2},\frac{\qb}{2})$
for 6-state QKD.
If the syndrome ($\qs={\tt Syn}\,x$) is sent in the clear, the right hand side of (\ref{refQKD}) acquires an extra $\sum_\qs$ outside
the trace norm and a factor $\qd_{\qs,{\tt Syn}(s\oplus a)}$ inside the trace norm;
the effect on (\ref{refQKD2}) is an extra factor $2^{n-k}$ under the square root;
while this alteration reduces the threshold value $\ell_{\rm max}$ by an amount $n-k$,
it has no effect on the rate since OTP'ing the syndrome would incur a penalty of exactly the same size.

\subsection{Achievable rate}
\label{sec:embedded_proof}

In the analysis we do not explicitly write down contributions from the 
authentication failure probability. 
It is implicit that each MAC adds a term $2^{-\ql}$
to the overall security parameter.

\begin{theorem}
\label{th:rate_embedded}
The {\tt KRUE} protocol satisfies the ENC, KR and UE
properties as defined in Section~\ref{sec:secdefattackmodel}
while achieving the following asymptotic rate,
\be
	r_{\rm 4state}=1-3h(\qb)
	\quad ; \quad
	r_{\rm 6state}=1-h(1-\fr{3\qb}{2},\fr{\qb}{2},\fr{\qb}{2},\fr{\qb}{2}) - h(\qb).
\ee
\end{theorem}
In other words, the achievable rate is worse than the QKD rate by a term $h(\qb)$.

\underline{\it Proof of Theorem~\ref{th:rate_embedded}:} 
Because of the inclusion of $n-k+\ql+1$ extra bits in the augmented message $m$,
the asymptotic rate of the protocol is $\ell_{\rm max}/n-h(\qb)$.
We need to determine the value of $\ell_{\rm max}$ for both the UE and KR property separately
and take the smaller of the two.

\underline{\bf Part 1}.
First we note that (\ref{diffUncl}) is the difference of two sub-normalised states
that both have trace equal to $P_{\rm corr}$.
This immediately yields the bound $\|\cE_{\rm UE}-\cF_{\rm UE}  \|_\diamond\leq P_{\rm corr}$.
Furthermore, from (\ref{diffUncl}) we get
\be
	\|\cE_{\rm UE}-\cF_{\rm UE}  \|_\diamond=  \!\!\!
	\EE_{mbzue} \fr1{2^n}\!\!\sum_{a}  
	\left\|  \EE_{st} \qr_{bst}^{\rm E}\qy_{st} \sum_p2^{\ell+n-k}\qd_{s\oplus a,p\|[{\rm Red}(p)\oplus e]} 
	[\qd_{\qF_u(p), m \oplus z}-\EE_{m'} \qd_{\qF_u(p), m' \oplus z}]
	\right\|_1
\label{part1diamond}
\ee
which resembles~(\ref{refQKD}).
The main difference is the $2^{n-k}\sum_p \qd_{s\oplus a,p\|[{\rm Red}(p)\oplus e]}$.
In the derivation as shown in Appendix~\ref{app:QKD}, upon doubling as in (\ref{QKDdoubling})
applying the $\EE_u$ then yields instead of $\qd_{ss'}$ the following expression,
\be
	(2^{n-k})^2 \sum_{pp'}\qd_{pp'}\qd_{s\oplus a,p||(e\oplus{\rm Red}p)}\qd_{s'\oplus a,p'||(e\oplus{\rm Red}p')}
	=
	(2^{n-k})^2\qd_{ss'}\qd_{e,(s\oplus a)[k+1:n] \oplus{\rm Red}((s\oplus a)[:k]) }.
\ee
The factor $(2^{n-k})^2\qd_{e,\cdots}$, together with the $\EE_e$ outside the trace norm,
together have the same effect as having the plaintext syndrome in the QKD derivation:
a factor $2^{n-k}$ under the square root in (\ref{refQKD2}).
Asymptotically this yields $\ell_{\rm max}^{\rm uncl,4state}=n-2nh(\qb)$
and
$\ell_{\rm max}^{\rm uncl,6state}=n-nh(1-\frac{3\qb}{2},\frac{\qb}{2},\frac{\qb}{2},\frac{\qb}{2})$.

\underline{\bf Part 2}.
First we note that (\ref{diffKR}) is the difference of two sub-normalised states
that both have trace equal to $P_{\rm corr}$.
This immediately yields the bound $\|\cE_{\rm KR}-\cF_{\rm KR}  \|_\diamond\leq P_{\rm corr}$.
Furthermore, from (\ref{diffKR}) we find
\be
	\| \cE_{\rm KR}-\cF_{\rm KR} \|_\diamond =
	\fr12\EE_{mbu}\frac1{2^{n+\ell}}\sum_{a\tilde z} \left\|  \EE_{st} \rho_{b s t}^{\rm E}
	\qy_{st}[2^\ell\qd_{\Phi_u((s\oplus a)[:k]), m \oplus \tilde z}  -1] \right\|_1.
\label{diffKR2}
\ee
This expression very closely resembles (\ref{refQKD}), with $\tilde z$ precisely playing the role of~$c$,
and the term $\EE_{m'}\qd_{c,m'\oplus\qF_u(a\oplus s)}$ replaced by the constant~`1'.
Carrying the `1' through steps
(\ref{QKDdoubling}) and further in Appendix~\ref{app:QKD} yields the same result as the QKD derivation,
except for one important difference:
the $(s+a)[:\!\!k]$ restriction to the first $k$ bits yields a modification of $\qd_{ss'}$
to the first $k$ bits only. 
In the end result the parameter $n$ is entirely replaced by~$k$.
Hence we obtain asymptotically $\ell_{\rm max}^{\rm KR,4state}=k-kh(\qb)$ $=n(1-h(\qb))^2$
and
$\ell_{\rm max}^{\rm KR,6state}$ $=k+kh(\qb)-kh(1-\frac{3\qb}{2},\frac{\qb}{2},\frac{\qb}{2},\frac{\qb}{2})$
$=n[1-h(\qb)][1+h(\qb)-h(1-\frac{3\qb}{2},\frac{\qb}{2},\frac{\qb}{2},\frac{\qb}{2})]$.

It is easily seen that $\ell_{\rm max}^{\rm UE}\leq \ell_{\rm max}^{\rm KR}$.
For brevity we use shorthand notation $h=h(\qb)$ and $H=h(1-\frac{3\qb}{2},\frac{\qb}{2},\frac{\qb}{2},\frac{\qb}{2})$,
noting that $H>h$ and $H<2h$.
For BB84 encoding we see  $\ell_{\rm max}^{\rm KR}/ \ell_{\rm max}^{\rm UE}=\frac{(1-h)^2}{1-2h}\geq 1$.
For 6-state we see 
$\ell_{\rm max}^{\rm KR}/ \ell_{\rm max}^{\rm UE}=\frac{(1-h)(1+h-H)}{1-H}$
$=\frac{1-H+h(H-h)}{1-H}\geq1$.
\hfill$\square$\\

\underline{Remark:} 
In the zero-noise case ($\qb=0$) there is no syndrome mask~$e$.
Then we have,
without inequalities, \\
 $\| \cE_{\rm UE} - \cF_{\rm UE} \|_\diamond = \| \cE_{\rm KR} - \cF_{\rm KR} \|_\diamond 
 = \| \cE_{\rm QKD} - \cF_{\rm QKD} \|_\diamond = \EE_{m u b z a} \| \EE_{st} \qth_{st} \rho^{\rm E}_{bst} 
 \big[ 2^\ell \qd_{\Phi_u(s\oplus a),m \oplus z} - 1\big]\|_1$, i.e.
the security of QKD immediately implies UE and KR. 

Also note that for $\qb=0$ we can invoke Lemma 2 to prove UE; 
we can view the keys $k_{\rm fb}$ and $k_{\rm OTP}$ as `external' to the proof, 
e.g. replace $k_{\rm fb}$ by the existence of an authenticated channel
and {\em spend} one bit to mask the feedback~$\qo$.

For $\qb>0$ we are not allowed to invoke Lemma~\ref{lemma:impliesUE}, since not all the key material is carried to the next round in
unmodified form: upon {\tt accept} the $e$ is updated.
The $e$ plays an integral role in the bounding of the diamond norm (\ref{part1diamond})
and cannot be moved outside that part of the proof.

\section{Combining protocols}
\label{sec:combine}

In the protocol as detailed in Section~\ref{sec:embedded},
the key $\tilde e_{\tt accept}$ is transported in an UE manner.
This is in a sense `overkill' since the attacker model assumes that long-term keys such as $\tilde e$
will leak eventually.
In this section we explore alternative ways of transporting $\tilde e_{\tt accept}$
which yield a better rate.
We consider only large-message asymptotics.

We introduce the following modified protocol, which we will refer to as {\tt KRUE}$^*$.
The message $m$ (Fig.\ref{fig:embedded}) now contains $\mu$ but not~$\qk$.
The $\qk$ is received in a different manner, but still together with the 
qubits of the {\tt KRUE}$^*$ scheme.
Ignoring the missing $\qk$,
the asymptotic rate of {\tt KRUE}$^*$ itself equals the QKD rate:
$r_4^{{\tt KRUE}^*}=1-2h(\qb)$ for 4-state encoding
and
$r_6^{{\tt KRUE}^*}=1-h(1-\frac{3\qb}{2},\frac{\qb}{2},\frac{\qb}{2},\frac{\qb}{2})$ for 6-state encoding.

\subsection{Combining {\tt KRUE}$^*$ with QKD}
\label{sec:plusQKD}


Here we consider {\tt KRUE}$^*$ where 
$\qk$ is received One-Time-Padded over the classical channel, and
the OTP key comes from running QKD\footnote{
In case of QKD failure, Alice and Bob try QKD again.
It does not matter if Eve observes the number of QKD failures.
}
before {\tt KRUE}$^*$.
Compared to our original scheme {\tt KRUE}, the net difference is:
\begin{itemize}[leftmargin=4mm,itemsep=0mm]
\item
Increased round complexity. 
Even assuming the most efficient form of QKD, Alice needs at least one extra pass to Bob.
The {\tt KRUE}$^*$ qubits cannot be sent in the same pass as the QKD qubits, since QKD requires Bob to
respond to Alice before the QKD key can be estabished. 
\item
Better rate. 
The asymptotic rate is $r_4^{{\rm QKD}+{\tt KRUE}^*}=\frac{[1-2h(\qb)]^2}{1-h(\qb)}$
for 4-state encoding and
$r_6^{{\rm QKD}+{\tt KRUE}^*}=\frac{[1-h(1-\frac{3\qb}{2},\frac{\qb}{2},\frac{\qb}{2},\frac{\qb}{2})]^2}
{1-h(1-\frac{3\qb}{2},\frac{\qb}{2},\frac{\qb}{2},\frac{\qb}{2})+h(\qb)}$
for 6-state. 
This can be seen as follows.
Let $\mu\in\bits^L$. 
Sending $\mu$ 
via {\tt KRUE}$^*$ needs
$n=L/r^{{\tt KRUE}^*}$ qubits. 
The size of the syndrome is $nh(\qb)$. 
Creating the OTP key using QKD takes $nh(\qb)/r^{\rm QKD} $ qubits.
The total number of qubits is 
$L(r^{\rm QKD}+h(\qb))/(r^{\rm QKD})^2$.
\end{itemize}

The combined scheme has the UE property regarding the message $\mu$.
First, by composability it is safe to use the QKD key in any manner.
Second, by the UE property of {\tt KRUE}$^*$,
$\mu$ is secure even if all keys, including everything contained in $\qk$,
eventually leak.

\vskip2mm


Interestingly, the rate $r_4^{{\rm QKD}+{\tt KRUE}^*}$ that we achieve here is twice the rate of QKD followed by 
Gottesman's Unclonable Encryption scheme \cite{uncl}.\footnote{
The rate for that combination is obtained as follows.
The UE step needs $n_{\rm UE}=L/[1-2h(\qb)]$ qubits.
Then $n_{\rm UE}$ bits of key need to be refreshed using QKD;
this takes $n_{\rm QKD}=n_{\rm UE}/[1-2h(\qb)]$ qubits.
The rate is $L/(n_{\rm UE}+n_{\rm QKD})$
$=\fr12\cdot\frac{[1-2h(\qb)]^2}{1-h(\qb)}$.
}



\subsection{Combining {\tt KRUE}$^*$ with QKR}
\label{sec:plusQKR}

Now Bob receives $\qk$ via the `Quantum Alice and Silent Bob' QKR scheme \cite{SilentBob}.
Again, in case of QKR failure Alice and Bob can just keep trying the QKR until it succeeds.
(This is the case because the only purpose is to transport random keys for the next round.)
The scheme of \cite{SilentBob} has the following properties,
(i) its asymptotic rate equals the QKD rate;
(ii) Alice sends only qubits and no classical communication. 

The advantage of using QKR over QKD is that {\em everything can be sent in the same pass}.
Hence the combination {\tt KRUE}$^*$+QKR achieves the same rate as 
{\tt KRUE}$^*$+QKD but has better round complexity (only one pass from Alice).

\begin{theorem}
\label{th:plusQKR}
Let $P_{\rm KR}$ be a $\qe_1$-KR scheme in which Alice makes one pass.
Let $P_{\rm UE}$ be a $\qe_2$-KR, $\qe_3$-UE scheme in which Alice makes one pass.
Let $Q$ be the composition of $P_{\rm KR}$ and $P_{\rm UE}$ such that
Alice sends her messages in parallel, and the message of $P_{\rm KR}$
is used as key material in $P_{\rm UE}$.
Then $Q$ is
$(\qe_1+\qe_2)$-KR, and it is $\qe_3$-UE with respect to the message of $P_{\rm UE}$.
\end{theorem}

\underline{\it Proof:}
We consider the EPR version of $Q$. 
Eve creates a state that can be written as $\qr^{\rm A_1 B_1 A_2 B_2 E}$,
where the labels `1' and `2' refer to the EPR pairs intended for $P_{\rm KR}$ and $P_{\rm UE}$ respectively,
and A,B refers to the EPR parts going to Alice and Bob.
As in Section~\ref{sec:secCPTP} we introduce different notation for the same CPTP map
depending on the property that we are looking at (KR or UE).
Thus we have CPTP maps $\cQ_{\rm 1KR}$, $\cQ_{\rm 1UE}$, $\cQ_{\rm 2KR}$, $\cQ_{\rm 2UE}$, with
\bea
	&& 
	(\cQ_{\rm 2KR}\circ \cQ_{\rm 1KR})(\qr^{\rm A_1 B_1 A_2 B_2 E})
	= \cQ_{\rm 2KR}(\qr^{M_1 \tilde K_1 T_1\rm  A_2 B_2 E})
	= \qr^{\tilde K_1 T_1 M_2 \tilde K_2 T_2 \rm E}
	\\ &&
	(\cQ_{\rm 2UE}^{\tt acc}\circ \cQ_{\rm 1UE})(\qr^{\rm A_1 B_1 A_2 B_2 E})
	= \cQ_{\rm 2UE}^{\tt acc}(\qr^{M_1 K_1\tilde K_1 T_1\rm A_2 B_2 E})
	= \qr_{[\qO=1]}^{M_1 K_1\tilde K_1 T_1 M_2 K_2\tilde K_2 T_2\rm E}. 
\eea
With respect to the KR property, the ideal functionality is
$\cQ_{\rm 2KR}^{\rm ideal}\circ \cQ_{\rm 1KR}^{\rm ideal}$.
With respect to UE the ideal functionality is as follows.
In case of {\tt reject} there are no requirements.
In case of {\tt accept} the $M_2$ is protected by $\cQ_{\rm 2UE}^{{\tt acc},{\rm ideal}}$ even if $\cQ_{\rm 1UE}$ does not behave ideally;
hence the ideal functionality is described by the mapping
$\cQ_{\rm 2UE}^{{\tt acc},{\rm ideal}}\circ \cQ_{\rm 1UE}$.
We have
\bea
	(\cQ_{\rm 2KR}^{\rm ideal}\circ \cQ_{\rm 1KR}^{\rm ideal})(\qr^{\rm A_1 B_1 A_2 B_2 E})
	&=& 
	\cQ_{\rm 2KR}^{\rm ideal}(\qr^{M_1 \tilde K_1}\otimes \qr^{T_1\rm  A_2 B_2 E})
	=
	\qr^{\tilde K_1 M_2 \tilde K_2}\otimes \qr^{ T_1  T_2 \rm E}
	\\
	(\cQ_{\rm 2UE}^{{\tt acc},{\rm ideal}}\circ \cQ_{\rm 1UE})(\qr^{\rm A_1 B_1 A_2 B_2 E})
	&=&
	\cQ_{\rm 2UE}^{{\tt acc},{\rm ideal}}(\qr^{M_1 K_1\tilde K_1 T_1\rm A_2 B_2 E})
	\nn\\ &=&
	\qr^{M_2}\otimes\qr_{[\qO=1]}^{M_1 K_1\tilde K_1 T_1 K_2\tilde K_2 T_2\rm E}.
\label{proof2ideal2}
\eea
It is given that
$\| \cQ_{\rm 1KR}-\cQ_{\rm 1KR}^{\rm ideal} \|_\diamond\leq \qe_1$, and
$\| \cQ_{\rm 2KR}-\cQ_{\rm 2KR}^{\rm ideal} \|_\diamond\leq \qe_2$, and
$\| \cQ_{\rm 2UE}-\cQ_{\rm 2UE}^{\rm ideal} \|_\diamond\leq \qe_3$.
The KR property of $Q$ follows trivially from
\be
	\Big\| \cQ_{\rm 2KR}\circ \cQ_{\rm 1KR} - \cQ_{\rm 2KR}^{\rm ideal}\circ \cQ_{\rm 1KR}^{\rm ideal} \Big\|_\diamond
	\leq
	\Big\|  \cQ_{\rm 1KR} -  \cQ_{\rm 1KR}^{\rm ideal} \Big\|_\diamond
	+\Big\|  \cQ_{\rm 2KR} -  \cQ_{\rm 2KR}^{\rm ideal} \Big\|_\diamond 
	\leq \qe_1 + \qe_2.
\ee
The UE property with regard to $M_2$ follows from
\be
	\Big\| \cQ_{\rm 2UE}\circ \cQ_{\rm 1UE} - \cQ_{\rm 2UE}^{\rm ideal}\circ \cQ_{\rm 1UE}  \Big\|_\diamond
	\leq
	\Big\| \cQ_{\rm 2UE} - \cQ_{\rm 2UE}^{\rm ideal}  \Big\|_\diamond
	\leq \qe_3.
\ee
\hfill$\square$

{\it Remark}.
It is possible to send the current-round $e$ via QKR instead of the next-round key $e'$.
This would make $e$ into a short-term variable instead of a long-term key,
and would make it possible to elegantly use Lemma~\ref{lemma:impliesUE} in the security proof.
However, it would also complicate the security analysis of the {\em combined} scheme.
We will not pursue this possibility here.


\section{Comparison to other schemes}
\label{sec:tradeoff}

We briefly comment on the round complexity
and the asymptotic rate of the protocols proposed in this paper as compared to other schemes.
The word `round complexity' here is not to be confused with the $N$ rounds in our protocol.
For a given message chunk $\mu_j$
we count {\em the number of times Alice has to send something}, 
and refer to this number as Alice's number of {\em passes}.
The rate is defined as the size of the message divided by the number of qubits.

We compare against other information-theoretically secure schemes which also do not use up\footnote{
Our schemes use up key material, but this is amortised over $N$ rounds. 
We neglect this expenditure for the purpose of the comparison.
}
key material,
\begin{itemize}[leftmargin=4mm,itemsep=0mm]
\item
{\bf QKD+OTP}.
Key establishment using Quantum Key Distribution, 
followed by One Time Pad classical encryption.
We consider efficient QKD with negligible waste of qubits \cite{LoChauArdehali2004} and the smallest possible number of 
communication rounds: only 2 passes by Alice.
\item
{\bf QKR}.
Qubit-wise prepare-and-measure Quantum Key Recycling as described in \cite{QKR_noise,SilentBob}.
Only a single pass by Alice is needed, since Alice and Bob already share key material.
\item
{\bf QKD+\cite{uncl}}.
Key establishment using QKD, followed by  Gottesman's Unclonable Encryption \cite{uncl}.
At least two passes by Alice are needed. 
\item 
{\bf QKR+\cite{uncl}}.
Key establishment using QKR, followed by  Gottesman's Unclonable Encryption. 
Only a single pass by Alice is needed when the two are performed in parallel \footnote{We don't give a proof for this combination as \cite{uncl} uses a different proof technique. Intuitively security is the same as the combination described in Section \ref{sec:plusQKR}}.
\end{itemize}

\begin{table}[h]
\begin{center}
{\small
\begin{tabular}{|l|c|c|c|}
\hline
&Alice &Asymptotic& \\
Protocol & \#passes & rate (4-state)  & Unclonability \\
\hline
QKD + OTP & 2 & $1-2h(\qb)$ & no\\
QKR \cite{QKR_noise,SilentBob} & 1 & $1-2h(\qb)$ & no\\
QKD + \cite{uncl} & 2 & $\frac12\cdot\frac{[1-2h(\qb)]^2}{1-h(\qb)}$  & yes\\
QKR + \cite{uncl} & 1& $\frac12\cdot\frac{[1-2h(\qb)]^2}{1-h(\qb)}$  & yes\\
{\tt KRUE} & 1 & $1-3h(\qb)$ & yes\\
QKD + {\tt KRUE}$^*$ & 2 & $ \frac{[1-2h(\qb)]^2}{1-h(\qb)}$ & yes\\
QKR + {\tt KRUE}$^*$ & 1 & $ \frac{[1-2h(\qb)]^2}{1-h(\qb)}$ & yes\\
\hline
\end{tabular}
}
\caption{\it Comparison of schemes that have no net expenditure of key material upon {\tt accept}.}
\label{tab:rates}
\end{center}
\vspace{-5mm}
\end{table}


\begin{figure}[h]
\begin{center}
\includegraphics[width=0.7\textwidth]{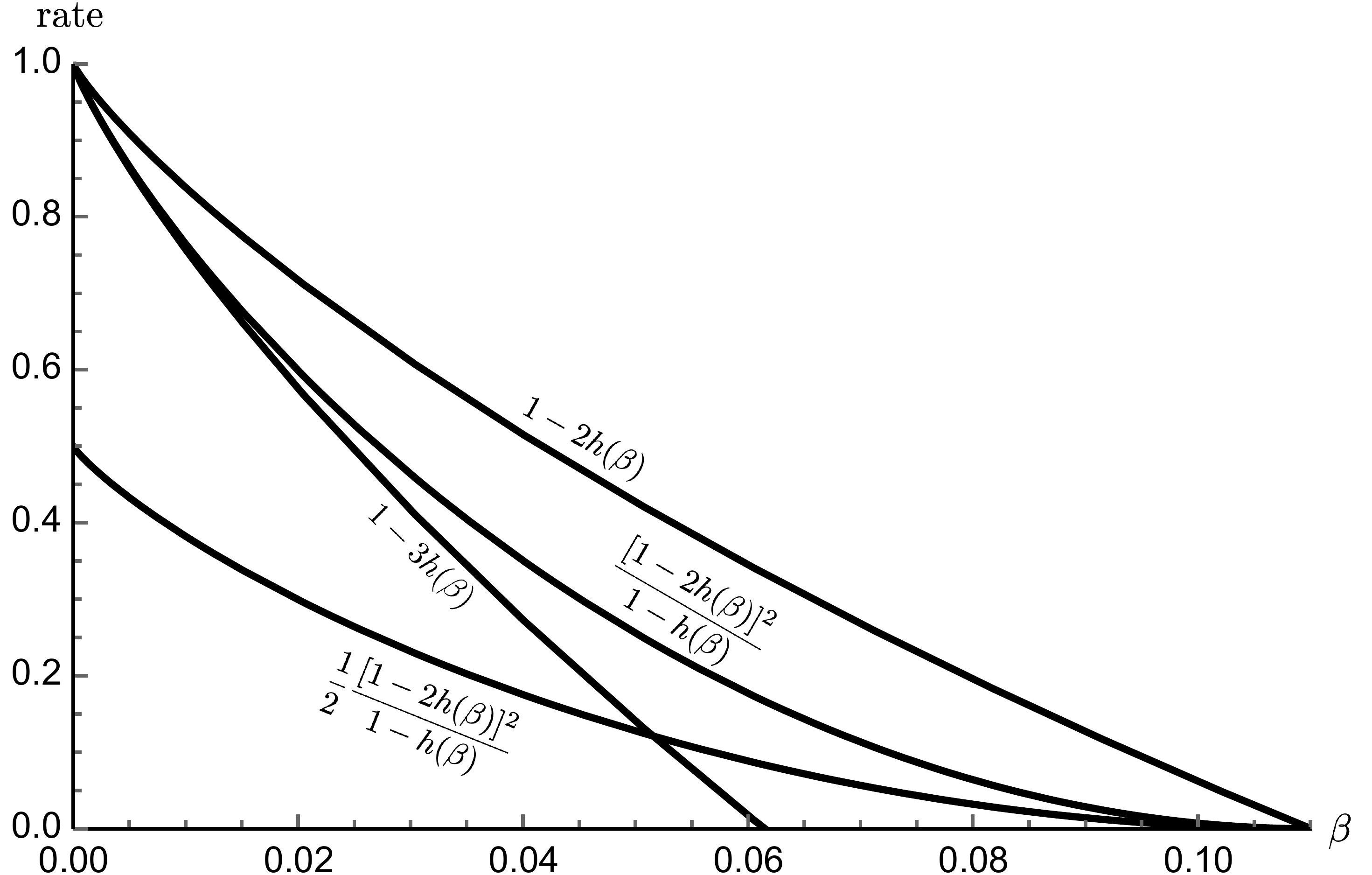}
\vspace{-3mm}
\caption{\it Asymptotic communication rates (4-state) as a function of the noise parameter~$\qb$.}
\label{fig:rates}
\end{center}
\end{figure}

The scheme properties are summarised in Table~\ref{tab:rates}, 
and the rates are plotted in Fig.\,\ref{fig:rates}. 
(We only show 4-state encoding.
The comparison holds qualitatively for 6-state encoding as well, but with slightly higher rates.)
QKR is an improvement over QKD in terms of round complexity, while achieving the same rate.
However, QKD and QKR over a noisy channel do not have the Unclonable Encryption property.

To our knowledge, the only existing scheme with an explicit proof of the UE property before our work
was Gottesman's construction \cite{uncl}.
(And thus ``QKD/QKR + \cite{uncl}'' was the only known way to have UE without net expenditure of key material.)
Our best performing scheme is QKR+{\tt KRUE}$^*$, with one pass from Alice and
double the rate of QKR + \cite{uncl}.
Our sub-optimal scheme {\tt KRUE} has a better rate than QKD/QKR + \cite{uncl}
at noise levels below $\qb\approx0.052$.

The above comparison does not contain the KR schemes {\cite{DBPS2014,FehrSalvail2017},
because \cite{DBPS2014} is defined only for the noiseless case $\qb=0$,
while \cite{FehrSalvail2017} 
has low rate ($\leq\fr13$) and limited noise tolerance.
Note that \cite{DBPS2014} has the UE property by Lemma~\ref{lemma:impliesUE},
and we suspect that \cite{FehrSalvail2017} satisfies a version of unclonability with a somewhat modified
definition that allows for a reduction of the min-entropy of some of the keys.
We believe that the QKR scheme \cite{QKR_noise} can be tweaked to have the UE property
by doing more privacy amplification; this would probably lead to the same rate as {\tt KRUE}.




We briefly comment on the key sizes.
The key material used in {\tt KRUE}
consists of the OTP $z\in \bits^\ell$, the hash seed $u \in \bits^k$, the basis choice $B \in \cB^n$, the redundancy mask $e\in \bits^{n-k}$, the authentication keys $k_{\rm MAC} \in \bits^\ql$, $k_{\rm fb} \in \bits^\ql$ and the OTP $k_{\rm OTP} \in \bits$. 
Counting only contributions proportional to $n$, the total size in bits is 
$\ell+n +n\log \cB+\cO(1)$. 
With $\ell\approx L+nh(\qb)$ and $n\approx L/[1-3h(\qb)]$ we can write the total size as
$L\frac{2+\log|\cB|-2h(\qb)}{1-3h(\qb)}+\cO(1)$.

The keys are expended over a block of $N$ rounds (or $\leq N$ in case of {\tt reject}).
If there are no {\tt reject}s, the `amortised' key expenditure
per round equals the above key size divided by $N$, which can be made much smaller than~$L$.

The key size of QKR+{\tt KRUE}$^*$ has a slightly different dependence on the noise parameter~$\qb$.
For the 4-state case, sending an $L$-bit message via {\tt KRUE}$^*$ requires 
$L+2n+\cO(1)$ bits of key, with $n\approx L/[1-2h(\qb)]$.
Sending $nh(\qb)$ bits via the QKR scheme \cite{SilentBob} takes a further $4\frac{nh(\qb)}{1-2h(\qb)}$ bits. 
This adds up to $L[1+\frac{2}{[1-2h(\qb)]^2}]+\cO(1)$ bits.

Gottesman's scheme has somewhat shorter keys,
total length $L\frac{2-h(\qb)}{1-2h(\qb)}+\cO(1)$,
but it needs to refresh $\approx L/[1-2h(\qb)]$ bits every round.

\section{Discussion}
\label{sec:discussion}

We have proven, 
in the proof framework developed by Renner et al.,
that quantum encryption can have Unclonability (as defined by Gottesman) as well as Key Recycling in the case of noisy channels. We achieve low communication complexity: there is no need for classical communication from Alice to Bob. 
The communication rate of {\tt KRUE} is $1-3h(\qb)$ for 4-state encoding 
and can be increased to $\frac{[1-2h(\qb)]^2}{1-h(\qb)}$ when combining our scheme with QKD or QKR.
Our scheme works by starting from QKR and
making the privacy amplification a step in the computation of the qubit payload.
Gottesman's construction \cite{uncl} does something very similar, and hence one might try
to construct a variant of our UE-KR protocol
that is closer to \cite{uncl}. 
This would have the advantage that there is no longer a seed $u$ that needs to be stored as part of the keys,
as \cite{uncl} employs ECC-based privacy amplification.
However, the proof technique that we use,
with its reliance on hash families, does not work for ECC-based privacy amplification.

Our best performing scheme, QKR+{\tt KRUE}$^*$, has a lower rate than QKD/QKR. 
(But interestingly the rate is positive on the same $\qb$-interval.) 
It is an open question whether
the rate decrease of UE schemes with respect to QKD is unavoidable.
The error-correction redundancy data has to be somehow protected;
this requirement does not exist in QKD.
Yet, the UE requirement makes it difficult to protect the redundancy, as long-term keys will leak eventually.
Perhaps an error-correcting scheme like \cite{DodisSmith2005}, which was used in \cite{FehrSalvail2017}, can help here.


\vskip2mm

Our protocols (temporarily) hide the {\tt accept}/{\tt reject} feedback bit~$\qo$.
This is a technicality that allows us to re-use~$b$ in un-altered form.
The alternative would be to send $\qo$ in the clear and then either (i) partially refresh $b$ as in \cite{QKR_noise},
or (ii) find a way to cope with a reduced entropy of $b$ as in \cite{FehrSalvail2017}.
Note that it is not realistic to hide a {\em large} accumulation of $\qo$-feedbacks from Eve.
Alice and Bob would have to act for a long time in a way that, to an external observer, does not depend
on the $\qo$s.
However, Eve may be able to observe e.g.~how often Alice and Bob have to engage in QKD to refill their key `reservoir',
which reveals the total number of {\tt reject}s.
For a {\em small} accumulation (e.g.~size $N$) we expect that it {\it is} realistic to hide the feedbacks temporarily. 

The downside associated with encoding a message directly into qubits 
is the vulnerability to erasures (particle loss) on the quantum channel.
Whereas QKD can just ignore erasures, 
in QKR they have to be compensated by the error-correcting code, which incurs
a serious rate penalty.


\vskip3mm

{\bf\Large Acknowledgements}\\
Part of this research was funded by NWO (CHIST-ERA project 651.002.003, ID\_IOT).

\newpage
\appendix
\section{QKD asymptotics}
\label{app:QKD}

We consider a QKD version that looks as much as possible like our protocol,
and apply Renner's proof technique to quickly derive bounds on the diamond norm.
For brevity we ignore message authentication tags and their failure probability,
since they do not affect the asymptotics.
We do not consider two-way postprocessing tricks like advantage distillation.
We refer to the resulting rates in this Appendix as the asymptotic rate of
QKD-with-one-way-postprocessing. 

\vskip2mm

\underline{QKD Protocol}.\\
Eve sends EPR pairs, in the singlet state.
Alice and Bob randomly choose measurement bases from the set~$\cB$,
perform their measurements,
and then publicly announce their basis choices.
They disregard all events where they chose different bases, and are left with $n$ bits.
Alice has measurement outcome $s\in\bits^n$, Bob has $t\in\bits^n$.
Alice generates random $x\in\bits^n$, $u\in\bits^n$. 
She computes a mask $a=s\oplus x$ and OTP $z=\qF_u(x)$.
She sends $a$ to Bob over an authenticated channel.
She also sends the syndrome $\qs={\tt Syn}(x)\in\bits^{n-k}$, either in the clear or OTP'ed.
(We will analyze both options.)

Bob computes $x'=t\oplus \bar a$ and tries to reconstruct $x$ from $x'$ and~$\qs$.
If he finds a $\hat x$ satisfying $|\hat x\oplus x'|\leq n\qb$ he sets $\qo=1$, otherwise $\qo=0$.
He sends $\qo$ to Alice.

In case $\qo=0$ Alice sets $c=\bot$.
In case $\qo=1$ she sets $c=m\oplus z$.
Alice sends $c,u$.
If $\qo=1$ Bob reconstructs $\hat z=\qF_u(\hat x)$ and $\hat m=c\oplus \hat z$.

\vskip2mm

\underline{Analysis in case of OTP'ed syndrome}.\\
Eve observes $b,u,a,c,\qo$ and holds a quantum state $\qr^{\rm E}_{bst}$ correlated to $b,s,t$. 
The message $m$ must be secure given Eve's information.
The output state of the QKD protocol is given by
\be
	\cE_{\rm QKD}(\qr^{\rm ABE})= \EE_{mbu}2^{-n}\sum_{ac\qo}\ket{mbuac\qo}\bra{mbuac\qo}\otimes\EE_{st}\qr^{\rm E}_{bst}
	\qd_{\qo,\qy_{st}}
	[\qo\qd_{c,m\oplus\qF_u(a\oplus s)}+\overline\qo \qd_{c\bot}].
\label{EQKD}
\ee
The idealized output state is obtained as $\EE_m\ket m\bra m\otimes \tr_{\!M}\cE_{\rm QKD}(\qr^{\rm ABE})$, which yields
\be
	\cF_{\rm QKD}(\qr^{\rm ABE})= \EE_{mbu}2^{-n}\sum_{ac\qo}\ket{mbuac\qo}\bra{mbuac\qo}\otimes\EE_{st}\qr^{\rm E}_{bst}
	\qd_{\qo,\qy_{st}}
	[\qo\EE_{m'}\qd_{c,m'\oplus\qF_u(a\oplus s)}+\overline\qo \qd_{c\bot}].
\ee
The difference is given by
\bea
	(\cE_{\rm QKD}-\cF_{\rm QKD})(\qr^{\rm ABE}) &=&
	\EE_{mbu}2^{-n}\sum_{ac}\ket{mbuac,\qo=1}\bra{mbuac,\qo=1}
	\nn\\ &&
	\otimes\EE_{st}\qr^{\rm E}_{bst}\qy_{st}
	[\qd_{c,m\oplus\qF_u(a\oplus s)} - \EE_{m'}\qd_{c,m'\oplus\qF_u(a\oplus s)}].
\label{QKDdiff}
\eea
This expression can be seen as the difference between two sub-normalized states which both
have norm $P_{\rm corr}$.
Hence an upper bound $\| \cE_{\rm QKD}-\cF_{\rm QKD} \|_\diamond \leq P_{\rm corr}$
immediately follows.
Furthermore, from (\ref{QKDdiff}) it follows that
\be
	\| \cE_{\rm QKD}-\cF_{\rm QKD} \|_\diamond \leq \fr12 \EE_{mbu}2^{-n-\ell}\sum_{ac}
	\left\|  \EE_{st}\qr^{\rm E}_{bst}\qy_{st}2^\ell
	[\qd_{c,m\oplus\qF_u(a\oplus s)} - \EE_{m'}\qd_{c,m'\oplus\qF_u(a\oplus s)}] \right\|_1.
\ee
Expanding the trace norm as $\| A \|_1=\tr\sqrt{A\dagg A}$ we write the right hand side as
\bea
\label{QKDdoubling}
	&& \!\!\!\!\!\!\!\!\!\!\!\!\!
	\fr12 \EE_{mbu}2^{-n-\ell}\sum_{ac}
	\\ && \!\!\!\!\!\!\!\!\!\!\!\!\!
	\tr\sqrt{
	\EE_{ss'tt'}\qy_{st}\qy_{s't'}\qr^{\rm E}_{bst} \qr^{\rm E}_{bs't'}2^{2\ell}
	[\qd_{\qF_u(a\oplus s),m\oplus c} - \EE_{m'}\qd_{\qF_u(a\oplus s),m'\oplus c}]
	[\qd_{\qF_u(a\oplus s'),m\oplus c} - \EE_{m''}\qd_{\qF_u(a\oplus s'),m''\oplus c}]
	}.
	\nn
\eea
Using Jensen's inequality for operators we `pull' $\EE_u$ and $\EE_m$ under the square root
and then make use of the pairwise-independent properties of $\qF_u$ when acted upon with $\EE_u$.
This yields
\bea
	2^{2\ell}\EE_{mu}[\qd_{\qF_u(a\oplus s),m\oplus c} \!-\! \EE_{m'}\qd_{\qF_u(a\oplus s),m'\oplus c}]
	[\qd_{\qF_u(a\oplus s'),m\oplus c} \!-\! \EE_{m''}\qd_{\qF_u(a\oplus s'),m''\oplus c}]
	\!\!\!\! &=& \!\!\!\!
	2^\ell\qd_{ss'}(1\!-\!\!\EE_{mm'}\qd_{mm'})
	\nn\\
	& < & 2^\ell \qd_{ss'}
\label{QKDdeltass}
\eea
which leads to
\be
	\| \cE_{\rm QKD}-\cF_{\rm QKD} \|_\diamond < \fr12\EE_b \tr\sqrt{ 2^\ell
	\EE_{ss'tt'}\qy_{st}\qy_{s't'}\qr^{\rm E}_{bst} \qr^{\rm E}_{bs't'}\qd_{ss'}
	}.
\ee
Next we use $\qy_{st}\leq 1$ and $\EE_t \qr^{\rm E}_{bst}=\qr^{\rm E}_{bs}$, yielding
$\| \cE_{\rm QKD}-\cF_{\rm QKD} \|_\diamond < \fr12\EE_b \tr\sqrt{ 2^\ell
	\EE_{ss'}\qr^{\rm E}_{bs} \qr^{\rm E}_{bs'}\qd_{ss'}  }$.
Combining the two obtained bounds gives
\be
	\| \cE_{\rm QKD}-\cF_{\rm QKD} \|_\diamond \leq 
	\min\Big(P_{\rm corr},
	\fr12\EE_b \tr\sqrt{ 2^\ell
	\EE_{ss'}\qr^{\rm E}_{bs} \qr^{\rm E}_{bs'}\qd_{ss'}  }
	\Big) .
\label{QKDdifffinal}
\ee
Using Post-selection, random Paulis and
smooth R\'{e}nyi entropy techniques, it has been shown \cite{RennerThesis,QKR_noise} that
the right hand side of (\ref{QKDdifffinal}) can be upper bounded as
$\propto\sqrt{2^{\ell-n+nh(\qb)}}$ for BB84 bases,
and as 
$\propto \sqrt{2^{\ell-n-nh(\qb)+nh(1-\fr32\qb,\fr\qb2,\fr\qb2,\fr\qb2)}}$
for 6-state QKD.

When $n$ is increased then either $P_{\rm corr}$ becomes exponentially small
(if Eve's noise $\qg$ exceeds $\qb$)
or (when $\qg\leq\qb$)
the expression under the square root becomes exponentially small,
provided $\ell$ is set smaller than some threshold value $\ell_{\rm max}$.
This threshold is given by 
$\ell_{\rm max}^{\rm BB84}=n-nh(\qb)$
and
$\ell_{\rm max}^{\rm 6state}=n+nh(\qb)-nh(1-\fr32\qb,\fr\qb2,\fr\qb2,\fr\qb2)$.
Taking into account the key expenditure for masking the syndrome ${\tt Syn}(x)$,
the asymptotic rate is $r=\ell_{\rm max}/n-h(\qb)$, i.e.
$r^{{\rm BB}84}=1-2h(\qb)$; $r^{6\rm state}=1-h(1-\fr32\qb,\fr\qb2,\fr\qb2,\fr\qb2)$.

\vskip2mm

\underline{Analysis in case of plaintext syndrome}\\
We indicate the differences w.r.t.~the analysis above.
Eq.\,(\ref{EQKD}) gains an extra part due to the syndrome $\qs$ and becomes
\be
	\cE_{\rm QKD}^{\rm plain}(\qr^{\rm ABE})  \!=\!\! 
	\EE_{mbu}\!\! 2^{-n}\!\sum_{ac\qs\qo}\ket{mbuac\qs\qo}\bra{mbuac\qs\qo}\otimes\EE_{st}\qr^{\rm E}_{bst}
	\qd_{\qo,\qy_{st}}\qd_{\qs,{\tt Syn}(a\oplus s)}
	[\qo\qd_{c,m\oplus\qF_u(a\oplus s)}+\overline\qo \qd_{c\bot}].
\label{EQKDplain}
\ee
The factor $\qd_{\qs,{\tt Syn}(a\oplus s)}$ is carried along untouched in the whole computation up to
(\ref{QKDdoubling}), where it gets doubled to $\qd_{\qs,{\tt Syn}(a\oplus s)}\qd_{\qs,{\tt Syn}(a\oplus s')}$.
However, the $\qd_{ss'}$ produced in (\ref{QKDdeltass}) undoes the doubling.
One extra step is needed. The sum $\sum_e$ is rewritten as $2^{n-k}\cdot\fr1{2^{n-k}}\sum_\qs$,
and Jensen's inequality is used, `pulling' the averaging operation $\fr1{2^{n-k}}\sum_\qs$ into the square root,
where it acts on $\qd_{\qs,{\tt Syn}(a\oplus s)}$, giving rise to a constant $2^{k-n}$.
\be
	\| \cE_{\rm QKD}^{\rm plain}-\cF_{\rm QKD}^{\rm plain} \|_\diamond \leq 
	\min\Big(P_{\rm corr},
	\fr12\EE_b \tr\sqrt{ 2^\ell 2^{n-k}
	\EE_{ss'}\qr^{\rm E}_{bs} \qr^{\rm E}_{bs'}\qd_{ss'}  }
	\Big).
\ee
The $\ell_{\rm max}$ is decreased by an amount $n-k$, but
the rate is exactly the same as before, since this time there is no key expenditure of $n-k$ bits
for encrypting the syndrome.


\newpage

\bibliographystyle{unsrt}
\bibliography{uncl_stripped}


\end{document}